# A search for monochromatic light towards the Galactic Centre

Geoffrey W. Marcy [ORCID],[1][*] Nathaniel K. Tellis[2] and Edward H. Wishnow[3]

[1]*Center for Space Laser Awareness, 3388 Petaluma Hill Rd, Santa Rosa, CA 95404, USA*
[2]*RocketCDL, Sausalito, CA 94965, USA*
[3]*Space Science Laboratory, University of California, Berkeley 94720 , USA*

**ABSTRACT**
A region 140 square degrees towards the Galactic Centre was searched for monochromatic optical light, both pulses shorter than 1 s and continuous emission. A novel instrument was constructed that obtains optical spectra of every point within 6 square deg every second, able to distinguish lasers from astrophysical sources. The system consists of a modified Schmidt telescope, a wedge prism over the 0.28-m aperture, and a fast CMOS camera with $9500 \times 6300$ pixels. During 2021, a total of 34 800 exposures were obtained and analysed for monochromatic sources, both subsecond pulses and continuous in time. No monochromatic light was found. A benchmark laser with a 10-m aperture and located 100 light years (ly) away would be detected if it had a power more than ~60 megawatt (MW) during 1 s, and from 1000 ly away, 6000 MW is required. This non-detection of optical lasers adds to previous optical SETI non-detections from more than 5000 nearby stars of all masses, from the Solar gravitational lens focal points of Alpha Centauri, and from all-sky searches for broadband optical pulses. These non-detections, along with those of broadband pulses, constitute a growing SETI desert in the optical domain.

## 1. INTRODUCTION

The Milky Way Galaxy and our local Solar system may contain technological entities, including civilizations, spacecraft, and data relay nodes that engage in communication using electromagnetic radiation (e.g. Cocconi & Morrison 1959; Bracewell 1960, 1973; Schwartz & Townes 1961; Freitas 1980; Maccone 2016; Gillon 2014; Hippke 2020, 2021; Gertz, 2018, 2021; Gertz & Marcy 2022). The transmitters and receivers may enable signals to travel thousands of light years. Alternatively, a network of nodes separated by less than a light year can overcome the inverse square attenuation of signals caused by the diffraction-induced spreading of the beam (Freitas, 1980, 1981; Gertz 2018; Gertz & Marcy 2022). Communication may occur with privacy, high bit rate, and minimal onboard energy by using the tight diffraction-limited beams of X-ray, ultraviolet, optical, or infrared lasers (Schwartz and Townes 1961; Freitas, 1980, 1981; Zuckerman 1985; Hippke 2018, 2021; Gertz & Marcy 2022). Laser beams are rapidly becoming a major mode of communication for satellites orbiting Earth (Hippke 2021c; Waterman 2022).

Even the most powerful lasers emit nearly monochromatic light, with wavelengths broadened somewhat by fundamental and engineering effects (Su et al. 2014; Naderi et al. 2016; Wang et al. 2020). We have previously searched for optical laser emission from individual stars using high-resolution spectra. We observed and examined spectra of more than 5000 normal stars of spectral type F, G, K, and M, including Sun-like stars, yielding no laser detections and no viable candidates (Reines & Marcy 2002; Tellis & Marcy 2015, 2017; Marcy 2021). A similar search for laser emission from more than 100 massive stars of spectral types O, B, and A has also revealed no viable candidates (Tellis et al., in preparation). These laser searches involved analysing high-resolution spectra, $\lambda/\Delta\lambda > 60\,000$, in the wavelength range $\lambda = 3600$–9500 Å, for monochromatic emission lines. The detection threshold of laser power was 50 kW to 10 megawatt (MW), assuming a diffraction-limited laser emitter consisting of a benchmark 10-m aperture. We also searched gravitational lens focal points (Marcy, Tellis, Wishnow 2022). No extraterrestrial lasers were found.

Searches for broad-band laser pulses (without spectroscopy) have been performed, e.g. Wright et al. (2001), Howard et al. (2004, 2007), Stone et al. (2005), Hanna et al. (2009), Abeysekara et al. (2016), and Villarroel et al. (2020, 2021). No definitive optical laser pulses were found. Next generation searches for optical pulses are in progress (Maire et al. 2020).

The Galactic Centre is an obvious direction to search for technology because of the large density of stars in that direction (e.g. Gajjar et al. 2021; Tremblay et al. 2022). Several searches for technology towards the Galactic Centre have been done at radio wavelengths (Gajjar et al. 2021; Tremblay et al. 2022). Here, we describe a search for monochromatic optical light towards the central $10 \times 14$ deg of the Milky Way. Laser light at optical wavelengths is unlikely to emerge from the Galactic Centre itself due to extinction by interstellar dust, except in a few narrow windows. However, the line-of-sight towards the Galactic Centre and Anti-Centre are special directions along which lasers may be pointing. For example, astronomers often point laser guide star beams towards the Galactic Centre to study the black hole there. Observers located inwards could look outwards towards the Galactic Anti-Centre to see our laser guide stars. Conversely, communicative entities located inwards of the Sun may purposely shoot laser beams outwards towards the Anti-Centre, knowing that curious astronomers will study the Galactic Centre. We could detect such lasers. As a bonus, the 10 × 14 deg search field obviously also pierces the local Galactic neighbourhood and the Solar system.

[*] E-mail: geoff.bnb@gmail.com

## 2. THE OBJECTIVE PRISM TELESCOPE

We used an objective prism Schmidt telescope operated by the Centre for Space Laser Awareness and described by Marcy, Tellis, and Wishnow (2022) and at www.spacelaserawareness.org. The telescope is a 'Celestron RASA-11' modified Schmidt design with an aperture of 0.28 m and a focal ratio of f/2.2. A 7-deg angle wedge prism of flint-type glass was commissioned to resolve the [O III] emission lines at 5007 and 4959 Å and to separate them from Balmer-$\beta$ emission. The goal was to identify ionized gas that might otherwise masquerade as laser emission. The telescope aperture, optics, and prism are similar to those used by Fleming, Cannon, and Pickering to produce the Henry Draper Catalog. Indeed, the spectra have a dispersion of ~450 Å mm$^{-1}$ near 4300 Å, similar to those from the Draper and Bache telescopes (Fleming et al. 1907; Pickering 1912; Cannon & Pickering 1922). Our only advance is the CMOS camera that brings 50x higher quantum efficiency, exposure times of 1 s, and a linear response of signal with intensity. Otherwise, this is century-old wide-field spectroscopy.

We employed a new CMOS sensor in the QHY600M camera purchased directly from QHYCCD in Beijing. We mounted the camera at the prime focus of the Schmidt telescope. The camera contains 9500 × 6300 pixels, each 3.7 μm, each with a quantum efficiency over 80 per cent between 500 and 800 nm, and each pixel having its own amplifier to allow 30 ms readout times of the entire image. The resulting field of view is 3.2 × 2.1 deg, and we operated with an exposure time of 1 sec. The system is sensitive to monochromatic pulses of optical light having pulse duration less than 1 s (see Marcy et al. 2022). The objective prism yields simultaneous spectra over the entire field. We do not use a diffraction grating because of the unfortunate 'zeroth-order' light having a PSF-shape from every star (and satellite glint) that vary in brightness due to seeing (Corbett et al. 2021; Nir et al. 2021). Finding a needle in the haystack does not benefit from thousands of mimics.

We performed multiple tests of read noise, dark noise, and linearity. The read noise is ~2 photons (RMS), the dark noise is less than 0.1 electron per second per pixel, and the response is linear within 0.3 per cent (and perhaps better) over a dynamic range of a few to 56 000 photo-electrons. This QHY600M CMOS sensor offers frame rates up to 4 frames s$^{-1}$ with a readout time of ~20 ms, enabling subsecond exposures to improve the contrast between light pulses having subsecond duration and the background 'noise' of stars, galaxies, and sky. From our observing station at Taylor Mountain in California, the sky produces ~40 photons per pixel during 1 s, coming mostly from city lights, giving a total 1$\sigma$ noise of ~7 photons per pixel per exposure.

Fig. 1 shows a typical image obtained with the objective prism telescope system and QHY600 CMOS camera. This image is the sum of 10 exposures each 1 s in duration, of 'Field 13' which has the Galactic Centre near the middle of the field (see Table 1). The image shows hundreds of stellar spectra oriented vertically, each spanning wavelengths 380–950 nm spread over 1200 pixels, with long wavelengths downward. For most stars, the red half of the spectrum appears brighter, highlighting 600 pixels in this rendering. In the 1 sec exposure, stars as faint as $V_{mag} = 13$ appear with signal-to-noise ratios (SNR) of ~10 per pixel. Stars brighter than $V_{mag} = 2.5$ saturate the sensor with > 56 000 photons per pixel. A monochromatic, spatially unresolved point source would appear as a two-dimensional 'dot' with a PSF shape. We judge the PSF by the width of the spectra in the spatial direction that is dominated by seeing and optical imperfections in the prism, yielding a PSF width of typically 6–7 arcsec, FWHM.

Figs 2 and 3 show spectrophotometry of Vega and NGC7027. Images with exposure times of 0.5 and 5 s were taken with our objective prism system. A simple reduction to one dimensional spectra was accomplished by summing the photons along the spatial width (Figs 2 and 3). The spectral lines identified exhibit a nonlinear wavelength scale due to the higher refractive index of the prism glass at shorter wavelengths. The optical spectrum spans nearly 1200 pixels, with the spectral and spatial resolution set by the PSF that has FWHM ~ 5.5 pixels, dominated by seeing and optical aberrations in the prism. The resulting spectral resolution varies from ~20 to 100 Å between 3800 and 9500 Å, respectively. The spectrum of Vega shows the Balmer lines up to H11, along with telluric lines. The spectrum of NGC7027 shows the usual emission lines from ionized gas at 10 000 K (e.g. Zhang & Li 2003), and the two [O III] at 4959 and 5007 Å are barely resolved. This modest spectral resolution allows stars, galaxies, ionized gas, asteroids, airplanes, and orbiting satellites to be identified and distinguished from other sources, notably non-astrophysical sources.

The spectrophotometry of Vega in Fig. 2 is given in photons per second per pixel detected with our objective prism system. The astronomical conventional photometric magnitude system is normalized to Vega at magnitude 0.0 in all UBVRI broad-band filters (Bessell 2005), allowing this spectrophotometry to map magnitude to photons per second per pixel. The monotonic decrease in photons detected for wavelengths shortwards of 5000 Å is due to decreasing quantum efficiency of the CMOS sensor and increasing dispersion (fewer Å per pixel). The key attribute of the reference spectra in Figs 2 and 3 is the spectral resolution that varies from 20 to 100 Å from 3800 to 9500 Å, which is sufficient to identify astrophysical objects, and hence provide an alert of anomalous, non-astrophysical objects, such as technological light sources.

Figs 2 and 3 allow scaling to fainter targets. For example, stars of magnitude $V = 10$ will yield ~80 photons per second per pixel at wavelengths of 5000 to 7000 Å. Noise from sky brightness is ~7 photons (rms) per pixel, making spectra of magnitude $V = 13$ stars comparable to sky brightness in exposure times of 1 s. With longer exposures, the persistent sources will acquire photons linearly with time while Poisson sky noise increases only as the square root of time, allowing sources slightly fainter than 13th magnitude to be detectable. Emission lines containing ~100 photons in the peak pixel will constitute ~10$\sigma$ detections.

## 3. OBSERVATIONS OF THE GALACTIC CENTRE

Fig. 4 shows the 25 fields we observed, each of angular size 3.2 × 2.1 deg, in the region towards the Galactic Centre. The entire mosaic is 14 × 10 deg, and the Galactic Centre is highlighted in bold at the centre. Each observation of a field consisted of 600 exposures, each 1 s in duration. We observed each field twice, except fields 12, 13, 14, 19, 20, 24, and 25 that we observed three times, as listed in

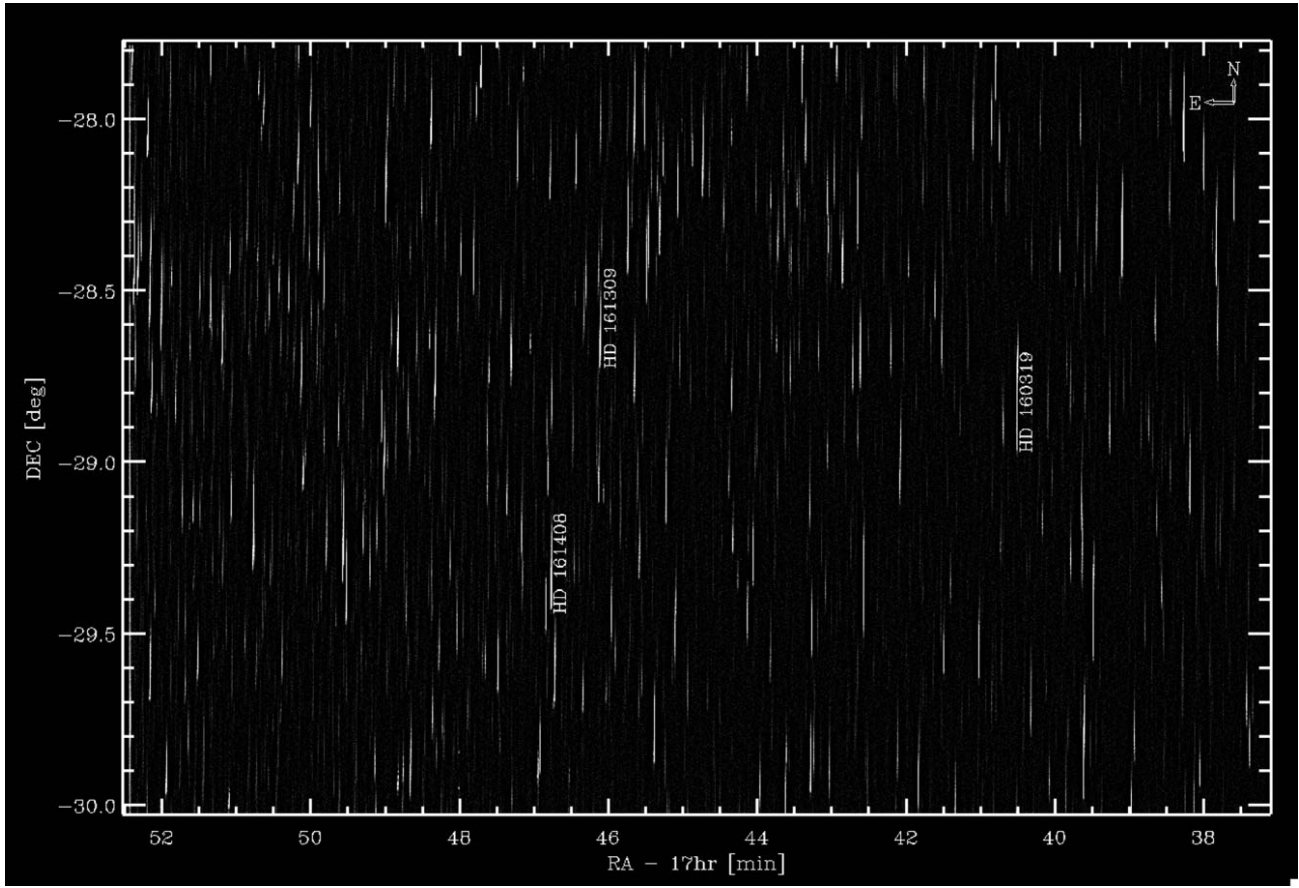

**Figure 1.** A typical objective prism image from the optical system with a field of view of 3.2 × 2.1 deg, 9500 × 6300 pixels, each subtending 1.3 arcsec on the sky. The stellar spectra span wavelengths 380–950 nm with longer wavelengths downwards. This image is centred on the Galactic Centre at RA = 17 h 45 m, Dec. = −29° 00′ with north up and east to the left. This is a co-add of 10 exposures, each 1 s duration. Spectra of three HD stars of magnitude $V \sim 7.5$ are labelled for reference. The other stellar spectra come from stars of magnitude 8–13. Monochromatic emission would appear as a PSF-shape dot.

Table 1. The fields overlap to provide both assurance of complete coverage of the region and security against algorithmic weakness at the edges of the field such as from poor background light assessment.

A set of 600 exposures will detect a variety of pulse durations and cadences, and miss others. Pulses having a duration shorter than 1 s will be detected if the number of photons is above the threshold for detection, roughly 400 photons, as quantified in Section 4.3. For each of the 25 fields, we obtained a second set of 600 exposures, separated by at least a few days, in an effort to detect pulse cadences that have low duty cycles. For example, a source may emit a pulse only once per day or week, in which cases the two sets of observations separated by a few days offer an increased chance of detecting at least one of them and perhaps a second occurrence. Clearly, additional observations would add to the detectable parameter space of pulse duration and cadence, albeit requiring increasing amounts of observing time.

## 4. THE SEARCH FOR MONOCHROMATIC PULSES

### 4.1 The image-difference search algorithm

We search for pulses of monochromatic emission, that appear as transient PSF-like 'dots' in the images, by employing a standard image-difference technique to search for sources that are consistent with the PSF shape (Bailey et al. 2007). The algorithm operates on a set of 600 exposures, each 1 s duration, of a given field (Table 1). The algorithm considers each exposure, one by one, to be a target image. For comparison, it gathers the average of six images, the three images taken before and the three images after the target image, to serve as a 'bookend' reference image. The algorithm simply subtracts the bookend reference image from the target image to yield the 'image difference', having pixel values near zero. As a first approximation, this image difference has all objects and background sky light removed. The non-zero residuals are due to Poisson noise of the arrival of photons and to the variations in atmospheric 'seeing' from image to image that compromises the quality of the subtraction of stellar spectra.

Any seeing changes within 1 s, along with Poisson noise, cause residuals in the difference image that are up to 10 per cent of the counts in the pixels of the original stellar spectra. We suppress these residuals due to seeing by performing a 50-pixel boxcar smoothing of the difference image in the direction of the spectra, and we subtract that smoothed version from the original difference image. The result is a difference image that has the residual stellar spectra removed further, including those caused by seeing variations. Narrow emission lines in the target image having widths under 10 pixels remain unaffected by the 50-pixel boxcar smoothing.

We construct a metric of the positive departures from zero in the new residual image by computing the Poisson noise (from the original number of photons) and then computing the ratio of the new residual image to the Poisson noise to yield a local SNR. Detectable emission lines will stand above unity in this SNR. The algorithm demands that a (first cut) potential monochromatic point source have reduced chi-square $< 2$, SNR $> 2.5$, and a minimum number of photons of 8 at the peak pixel as the criteria that define individual 'pixels of interest.'

**Table 1.** ields observed during 2021 August and September.

| Field # | RA | Dec. | Dates |
|---|---|---|---|
| 1 | 17h 19m | −33° 00' | 8–27, 9–07 |
| 2 | 17h 19m | −31° 00' | 8–27, 9–07 |
| 3 | 17h 19m | −29° 00' | 8–31, 9–07 |
| 4 | 17h 19m | −27° 00' | 9–01, 9–07 |
| 5 | 17h 19m | −25° 00' | 9–03, 9–07 |
| 6 | 17h 32m | −33° 00' | 8–27, 9–08 |
| 7 | 17h 32m | −31° 00' | 8–27, 9–08 |
| 8 | 17h 32m | −29° 00' | 8–31, 9–08 |
| 9 | 17h 32m | −27° 00' | 9–01, 9–08 |
| 10 | 17h 32m | −25° 00' | 9–04, 9–08 |
| 11 | 17h 45m | −33° 00' | 8–27, 9–07 |
| 12 | 17h 45m | −31° 00' | 8–27, 9–07, 09–13 |
| 13 | 17h 45m | −29° 00' | 8–31, 9–04, 09–11 |
| 14 | 17h 45m | −27° 00' | 9–01, 9–07, 09–13 |
| 15 | 17h 45m | −25° 00' | 9–04, 9–08 |
| 16 | 17h 58m | −33° 00' | 8–31, 9–11 |
| 17 | 17h 58m | −31° 00' | 8–31, 9–11 |
| 18 | 17h 58m | −29° 00' | 9–01, 9–11 |
| 19 | 17h 58m | −27° 00' | 9–01, 9–08, 09–13 |
| 20 | 17h 58m | −25° 00' | 9–04, 9–04, 09–13 |
| 21 | 18h 11m | −33° 00' | 8–31, 9–04 |
| 22 | 18h 11m | −31° 00' | 8–31, 9–11 |
| 23 | 18h 11m | −29° 00' | 9–01, 9–11 |
| 24 | 18h 11m | −27° 00' | 9–04, 9–08, 9–13 |
| 25 | 18h 11m | −25° 00' | 8–27, 9–04, 9–11, 9–13 |

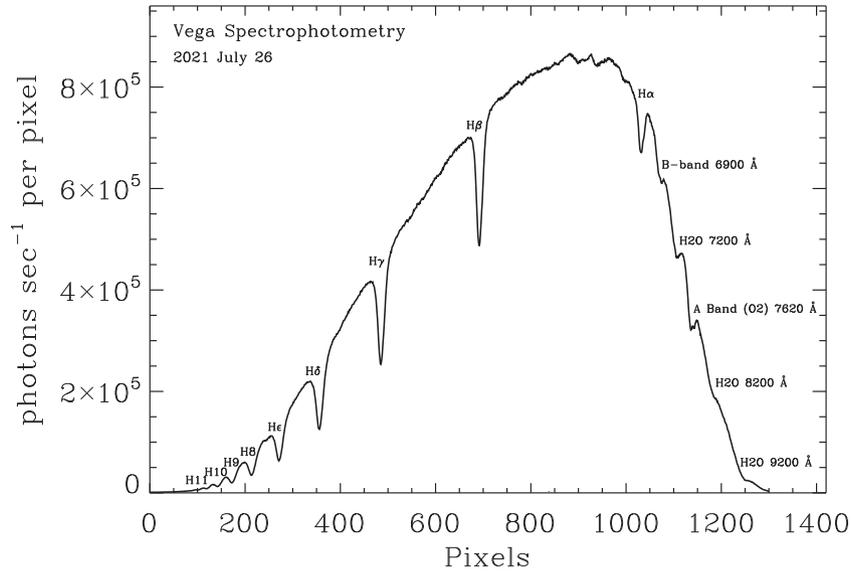

**Figure 2.** Spectrophotometry of Vega in photons per second per pixel detected with the objective prism 0.28-m RASA Schmidt telescope and its QHY600M CMOS camera. Wavelength increases to the right, with absorption lines labelled for reference. Vega is magnitude 0.0 in all UBVRI broad-band filters, allowing this spectrophotometry to map magnitude to photons per sec per pixel. The decrease in photons detected shortwards of 5000 Å (near H $\beta$) is due to decreasing quantum efficiency and increasing dispersion (less Å per pixel). The spectral resolution varies from 20 to 100 Å from 3800 to 9500 Å, respectively.

For each 'pixel of interest', the algorithm computes a final chi-square statistic and the RMS of the difference between the candidate point source and the measured instantaneous PSF (from the spatial profile) from that same image, both being normalized to the peak. We retain all monochromatic point sources that contain at least eight photons in the peak pixel and that have a profile compared to the measured PSF that yields reduced chi-square less than 1.8 and RMS less than 45 per cent of the peak. The smooth PSF with FWHM ∼ 5.5 pixels provides a robust criterion for identification of monochromatic point sources by the image-difference algorithm. Cosmic-ray muons are immediately rejected, as they affect only a few neighboring pixels. These detection criteria were painstakingly determined so that zero false positives would emerge due to noise fluctuations in the images, i.e. from sky photons or poor subtraction of stellar spectra. Indeed, we encountered no false positives in this search. The monochromatic point sources identified by the algorithm are catalogued for further analysis, notably the presence of other emission lines, as described in Section 4.2.

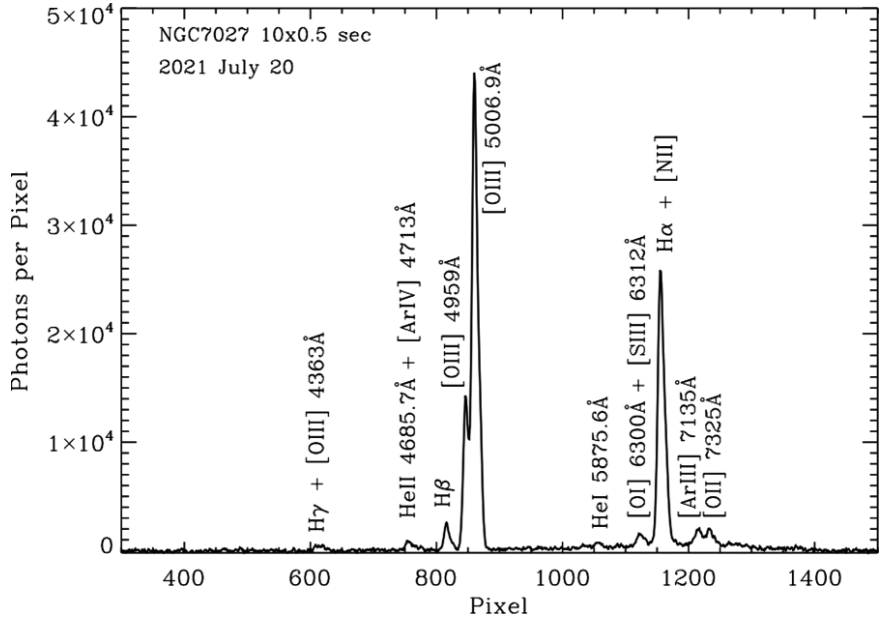

**Figure 3.** A spectrum of the planetary nebula NGC7027, obtained in 5 s, to test the discrimination of astrophysical line-emitting sources from technological monochromatic sources. Prominent emission lines from ionized gas are resolved from each other and identified here, including Balmer lines, [O III] 5007 Å and 4959 Å, [O I], He I, and He II. The spectral resolution is ~20 Å (in the blue) to 100 Å (in the red), enabling the discrimination of technological monochromatic sources from astrophysical sources.

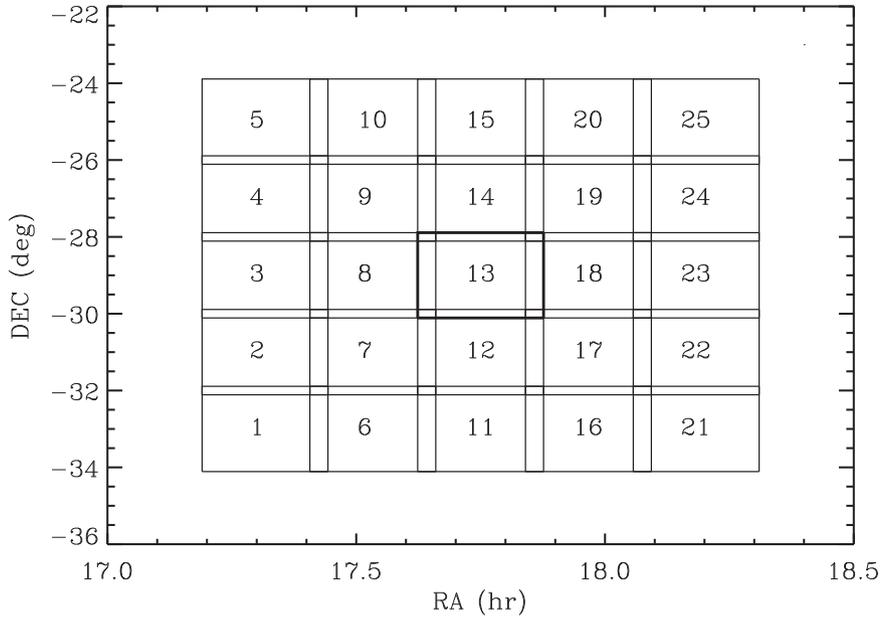

**Figure 4.** The Galactic Centre region observed as an overlapping mosaic of 25 fields, each 3.2 × 2.1 deg. The mosaic is 14 × 10 deg centred on RA = 17h45m Dec. = −29° 00 for which the central field (#13) is shown bold. Each field was observed at least twice with 600 exposures, 1-s each, using the objective prism optical system. Each exposure gives a spectrum of each point in space, able to reveal sub-second or continuous monochromatic optical pulses.

This image-difference result contains any monochromatic point sources that were present in the target image but not present (or only weakly present) in the average of the 6 'bookend' images taken before and after. Fig. 5 shows a demonstration of this process for a representative case, with a synthetic monochromatic sub-second pulse of light injected into the 'target' image (the 4th image). The top panel of Fig. 5 shows the time series of 7 consecutive raw images, each a 1 s exposure, shown zoomed to 1000 × 1000 pixels. The 4th (middle) image is the 'target' image with an injected synthetic monochromatic point source (barely visible) in the centre. The entire target image (9500 × 6300 pixels) was examined blindly by the image-difference algorithm to search for point sources. The middle row shows the result for the target (4th) image after subtracting the 6 'bookend' images taken before and after. The 4th image of the middle row shows the stellar spectra are effectively removed, allowing a monochromatic point source to be located in between them, and the image subtraction algorithm suppresses both stars and sky and all other persistent sources. The algorithm demands that the candidate point source must have a 2D shape consistent with the instantaneous point spread function (PSF), as measured by the spatial profile of the stellar spectra determined by cross-correlation. For every 1 s exposure, the algorithm measures the FWHM of the spatial profile of stellar spectra, commonly 5–6 pixels (6–8 arcsec), caused by seeing and optical aberrations in the prism.

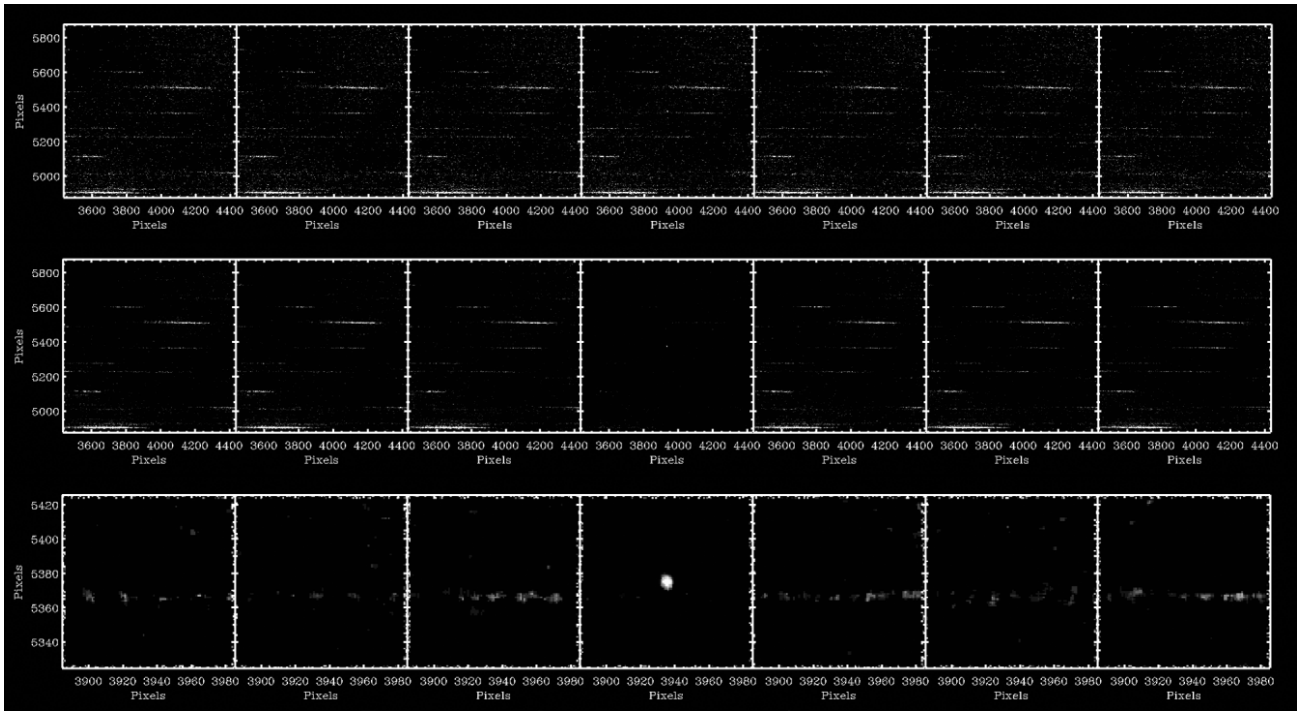

**Figure 5.** Demonstration of the image-difference algorithm. Top row: 7 consecutive raw images (1 s exposures), zoomed at 1000 × 1000 pixels, with a synthetic monochromatic point source injected at the centre of the 4th image (barely visible). Middle row: The same images, but the 4th image has the average of the other 6 images subtracted. Bottom row: Same as middle row, but with a zoom on 100 × 100 pixels that reveals, by eye, the injected synthetic monochromatic point source.

The PSF width varies by 15 per cent (rms) within a given image, as measured by examining the widths of stellar spectra at all regions of images. The search algorithm must accommodate that PSF variation. To find all candidate emission lines despite variations in the PSF across the image, the algorithm runs five trial searches that employ a range of assumed PSF widths, specifically the measured width (i.e. nominal for that exposure) and also 10 per cent and 5 per cent smaller and 5 per cent and 10 per cent larger than the measured PSF. In addition, the chi-square criterion accepts all dots that yield a reduced chi-square statistic up to 1.8 (with a perfect fit being 1.0). This approach implies that the discrepancies between an emission source (dot) and the measured PSF may be up to 20 per cent and still be retained as a candidate monochromatic emission. This flexibility in the PSF and chi-square criterion allows nearly all the emission dots to be detected, including PSF variability. The laser detectability may be slightly poorer at the extremes of wavelength.

This image-difference algorithm detects sub-second pulses of monochromatic emission pulses only if the emission is absent in the three images before and after. If, instead, the cadence of pulses is more frequent than 1 pulse per second, the image-difference algorithm will likely fail to detect a pulse due to cancellation of the pulse in the difference between the target image and the bookend reference image. In contrast, cadences slower than ∼1 pulse per second will yield individual frames with one pulse more intense than the average of the 3 + 3 frames taken before and after, making the pulse detectable with the difference algorithm.

Remarkably, point sources of monochromatic light that are *continuous in time* (e.g. continuous lasers) are nonetheless detected by the image-difference algorithm due to seeing changes on sub-second time scales. Scintillation causes steady sources of monochromatic light to vary in measured brightness by 10–20 per cent (RMS) in the exposures, for the same reason that broadband photometry is degraded by turbulence in the Earth's atmosphere. The image-difference algorithm detects those scintillation variations as if they were transient 'light pulses.' Thus, the natural and constant emission lines from planetary nebulae, magnetically active M dwarfs, and Wolf–Rayet stars invariably exhibit apparent 'transient' monochromatic light according to the image-difference algorithm simply due to seeing variations (see Section 4.2). In summary, monochromatic point sources that either last less than 1 s but have a slow cadence or are continuous in time can be detected by the image difference algorithm.

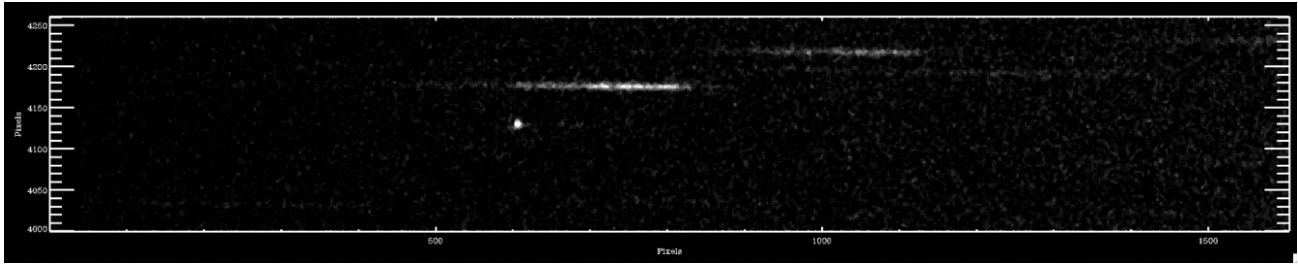

**Figure 6.** A monochromatic point source found in a 1-s exposure of field #12, locatied 2 deg south of the Galactic Centre. This emission was found using the automated difference-image and point-source search algorithm. This emission appears in all 1800 exposures of this field, spanning 41 d of observations. Fluctuations in seeing caused this persistent emission to be temporarily elevated above threshold in the difference image. This is a monochromatic object of interest – a candidate laser pulse worthy of further assessment.

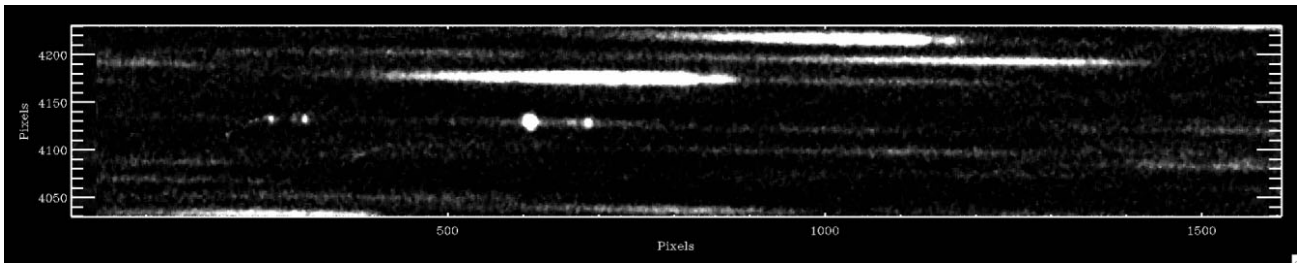

**Figure 7.** Co-added 100 × 1 s exposures of the candidate in Fig. 6. Multiple emission lines appear, two being redwards (right) of the original emission and at least three bluewards. The multiple lines suggest the source is astrophysical. Also visible are the spectra of a dozen stars, the faintest being $V_{\mathrm{mag}} \sim 15$, in this subimage.

### 4.2 Detected monochromatic objects of interest

Using the search algorithm described above, all 58 sets of 600 exposures towards the Galactic Centre region were searched. The explicit goal is to discover monochromatic point sources having a duration of less than 1 s. A fortuitous benefit of seeing variations is that continuous monochromatic emission, i.e. lasting more than ~10 s, is also identified by the search algorithm for subsecond pulses, as we describe below. All 25 regions within the 14 × 10 deg field towards the Galactic Centre were searched, as shown in Fig. 4 and Table 1. The image-difference-image algorithm described in Section 4.1 yielded a list of PSF-shaped 'monochromatic objects of interest,' i.e. candidates, that satisfied the criteria for monochromatic subsecond pulses. Several dozen monochromatic objects of interest emerged from the search, and each of them required visual inspection and assessment. We describe the objects of interest here.

One type of monochromatic object of interest is shown in Fig. 6. The image-difference analysis found a PSF-shaped monochromatic source in one 'target image' relative to the 6 adjacent 'bookend' reference images. This isolated 'dot' in one image relative to adjacent exposures is exactly what is expected from a monochromatic pulse, such as from subsecond emission from a distant, spatially unresolved laser.

Examination of the exposures taken before and after the target image shows they also have the monochromatic emission, albeit fainter. Thus, this is persistent monochromatic emission. We test for the presence of other emission lines by co-adding 100 exposures, each 1-s, as shown in Fig. 7. Indeed, that co-added image reveals other emission lines.

Fig. 8 shows the co-added 100 exposures collapsed in the spatial direction to create a 1D spectrum. That spectrum shows this object of interest is a planetary nebula with strong H α emission. Seeing variations during the 1-s exposures can increase the intensity of the H α emission line above the intensities of the exposures taken just before and after, fooling the image-difference analysis into detecting a 'pulse' of monochromatic emission. Fig. 8 shows the identification of the emission lines, all being the usual lines found in planetary nebulae. Several other planetary nebulae were 'discovered' in the images within the 14 deg x 10 deg surveyed here towards the Galactic Centre. All are well-known planetary nebulae that have been published and do not merit further analysis here.

A second type of monochromatic object of interest 'discovered' in our data by the image-difference algorithm is shown in Figs 9 and 10. The algorithm identified strong emission in one target image that stood out relative to the 6 adjacent 'bookend' reference images. Co-addition of 100 images revealed the spectrum of a magnetically active M-dwarf star of spectra type dMe. The suffix 'e' denotes H α emission, often accompanied by H β emission, and indicates strong magnetic fields and flares on the star's surface. Variable seeing fooled the image-difference analysis as the detected emission brightened momentarily. One must reject this monochromatic object of interest as merely a common magnetic, low-mass star.

The difference-image algorithm also 'discovered' two monochromatic objects of interest that, upon co-addition of images, turned out to be Wolf–Rayet stars. Their emission lines vary in intensity due to seeing changes, mimicking pulses. One example found in field #16 is shown in Fig. 11. This is the known Wolf–Rayet star, WR 103 = HD164270, with $V_{\mathrm{mag}} = 8.74$ (RA = 18 01 43.1, Dec. = −32 42 55, eq.2000). Our spectrum is consistent with that published by Williams et al. (2015), albeit at lower resolution. The emission lines are mostly from multiply-ionized carbon and also neutral and ionized helium.

One other type of candidate found by the image-difference algorithm is shown in Figs 12 and 13. The 'movie' frames in Fig. 12 show the seven raw 1-s images, magnified around the 'candidate' that was identified by the algorithm in the 4th frame. All seven images show an object that is apparently moving a few hundred pixels each second, from upper left to lower right. The change in intensity at a given location in the 4th frame triggered the algorithm. We interpret this sequence of images as caused by an airplane that is moving down and to the right.

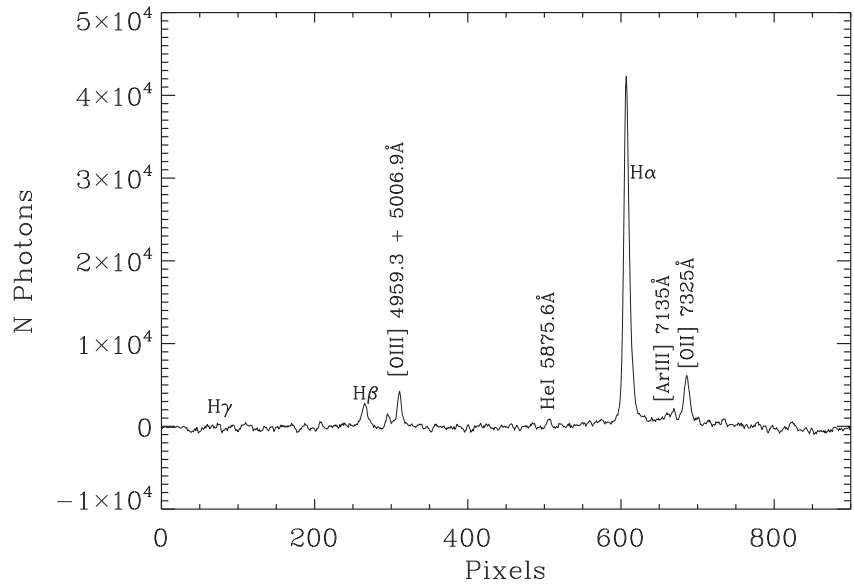

**Figure 8.** A 1D 'mash' of the emission lines from Fig. 7, with lines identified, indicating the source is a planetary nebula. See Fig. 3 for comparison with the planetary nebula NGC7027.

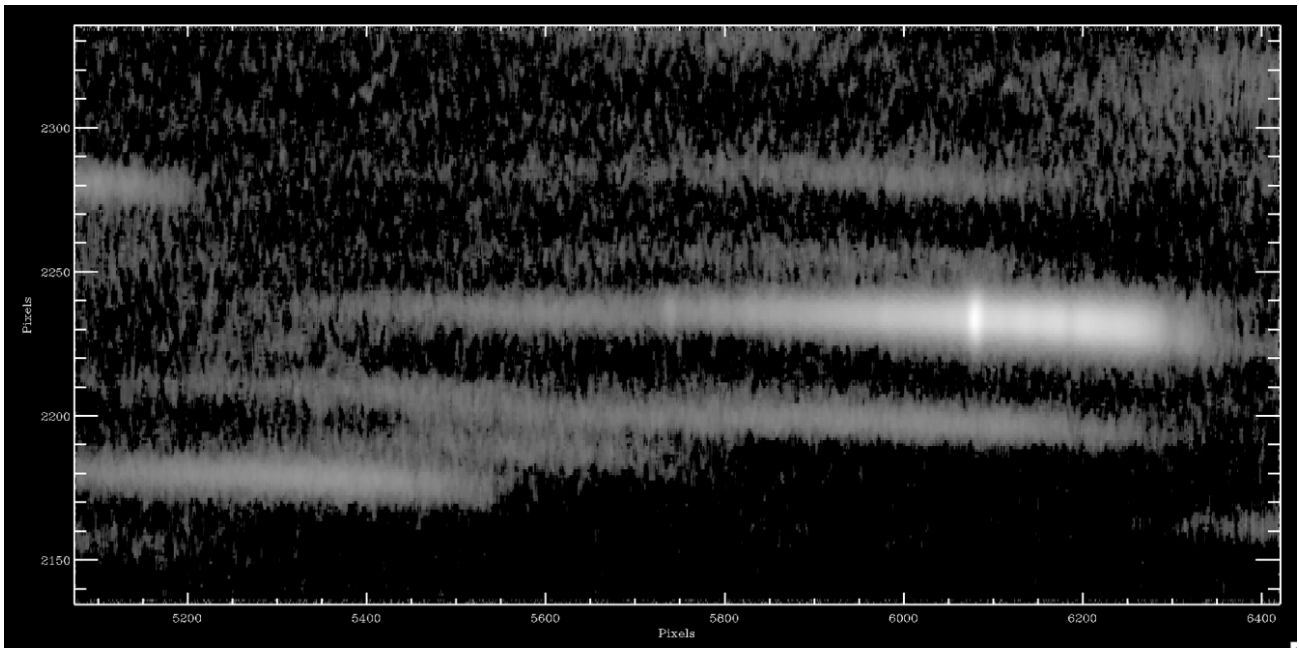

**Figure 9.** A representative magnetically active M dwarf of spectral type dMe found automatically in field #14 with the image-difference analysis of the objective prism 1-s exposures. This is a co-addition of 100 images. Seeing variations cause H $\alpha$ (pixel 6075) to be momentarily more intense than in the adjacent exposures. H $\beta$ is also in emission (pixel 5740), and the continuum is red, consistent with an M dwarf star.

The broad diagonal smear is a red-white light that is on all the time. That left-right horizontal extent of ~600 pixels of the smear gives the instantaneous spectrum, dominated by yellow and red light. The diagonal extent of the smear is caused by the motion during the 1 s exposure. The upper, narrow horizontal streak is a red light that pulses on for only ~0.05 s. The horizontal extent is 500 pixels long and confined to the red. Its narrowness indicates it was lasted for only 1/20 of the exposure time of 1 s. The longer horizontal streak is a broad-band light that includes blue, green, yellow, and red light, but its narrowness shows it, too, pulsed on for only ~0.05 s. Both the blinking red and broad-band lights pulsed with duration only 0.05 s, which keeps them narrow in the diagonal direction. The red and broad-band pulsing lights have a cadence of 1.2 s, allowing them to appear in most of the succession of frames. This collection of moving light sources demonstrates the value of the spectroscopic and temporal resolution. Moving sources offer temporal resolution of less than 1 s, in addition to spectral resolution of $\lambda/\Delta\lambda \sim 100$. For example, satellites in low earth orbit that emit laser light will reveal their velocity and both the wavelength and pulse cadence of their laser communication, including pulse sequences.

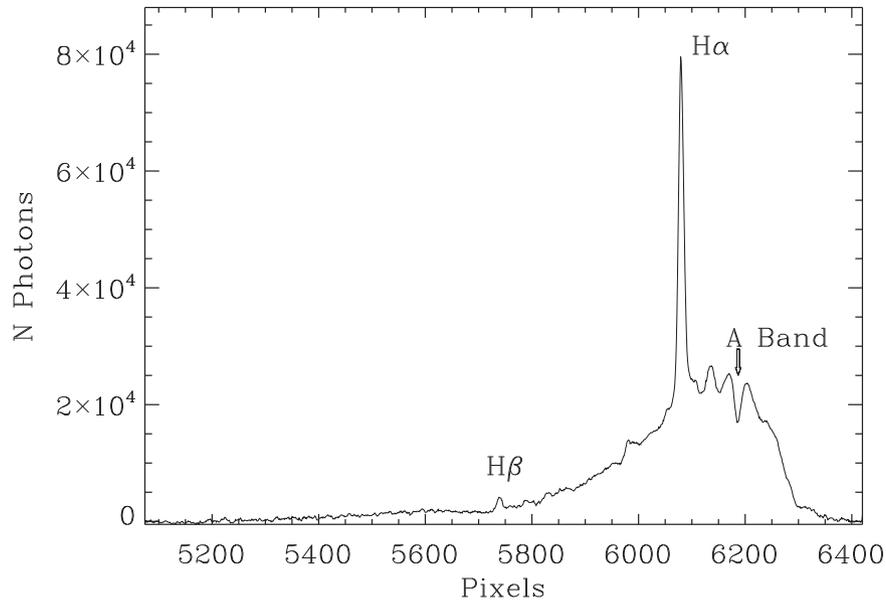

**Figure 10.** A 1D spectrum of the magnetically active M dwarf 'discovered' by the image-difference analysis, shown in Fig. 9. Emission is apparent at H $\alpha$ (pixel 6075) and H $\beta$ (pixel 5740), and the spectral energy distribution is red, consistent with an M dwarf, ruling out technological origin.

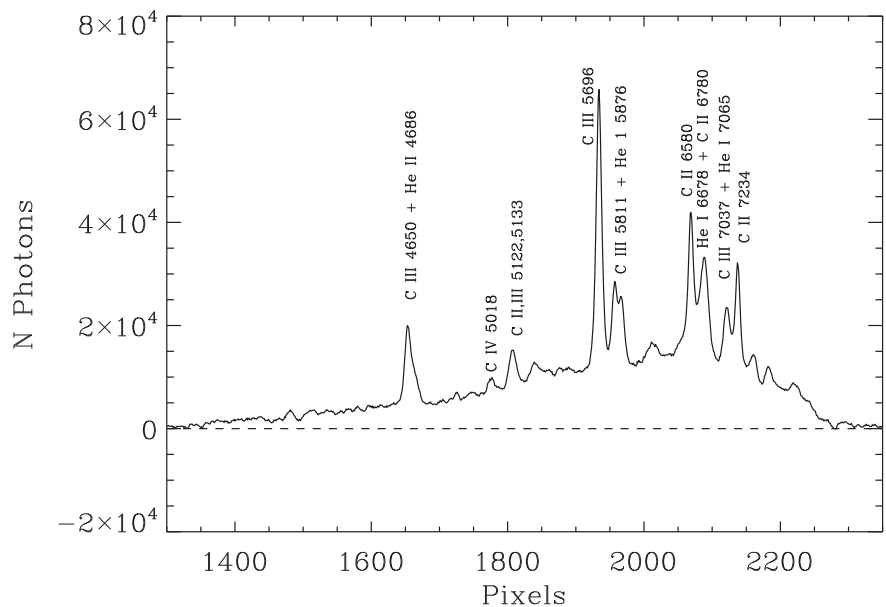

**Figure 11.** An emission-line star 'discovered' in the spectral imaging data using the automated difference-image algorithm. Seeing variations affect the apparent intensity of the emission lines, triggering a false alarm of a monochromatic pulse. This is the known Wolf–Rayet star, WR 103 = HD 164 270 ($V_{\text{mag}} = 8.74$). The plot shows the sum of 100 exposures, each 1 s.

Fig. 13 shows a magnified view of the spectrum of the broadband blinking light that pulses every 1.2 s with pulse duration of ~0.05 s. The spectrum shows structure with emission lines at the red end (far right). Such emission in the near-IR is consistent with the lamp on the aircraft being Krypton, Xenon, or Argon. Indeed, Xenon is commonly used for aircraft lights, and it has emission lines in the near-IR. This 0.05 s flashing broadband light with emission lines demonstrates the spectral, spatial, and temporal resolution of this optical system. Technological sources moving across the sky, such as unknown satellites, would be subject to similar analysis.

In summary, the algorithm performed a search of the multiple exposures of the 14 × 10 deg field towards the Galactic Centre, yielding several dozen monochromatic objects of interest. None of them were actually pulses of monochromatic light. Instead, all of them were either astrophysical objects with a strong emission line that varied due to seeing changes, or aircraft with flashing lights with duration ~0.05 s (including emission lines). We found no point-sources with monochromatic emission, pulsing or otherwise. None were plausibly extraterrestrial lasers.

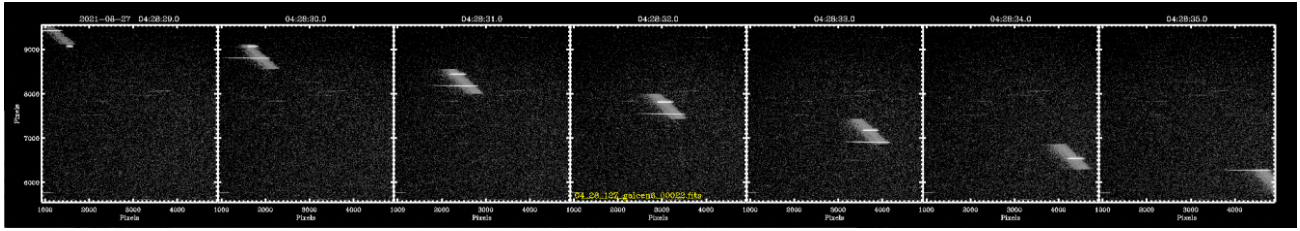

**Figure 12.** A transient source found towards the Galactic Centre. Seven consecutive frames are shown, each a 1 s exposure, with increasing wavelength to the right, and north to the left. The source moves diagonally from upper left to lower right. Three components of light appear. One source has a continuous spectrum, and is 'on' all the time. A second source spans only yellow and red wavelengths covering only 400 pixels, and it flashes every 1.2 s. A third source flashes every 1.2 s and exhibits all wavelengths from 400 to 900 nm (1000 pixels long). These are probably the lights from an airplane, one continuous and two flashing on the tips of the wings.

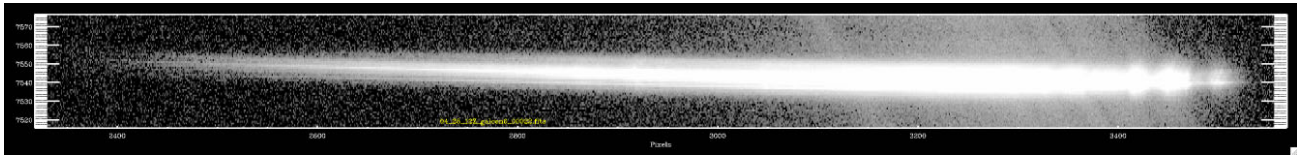

**Figure 13.** Magnified view of the spectrum of the broadband green point source in Fig. 12 that spans 1000 pixels, i.e. nearly all wavelengths. It is mostly a continuous spectrum, but also has emission structure in the near-IR region at the far right. This wide-band strobing light must contain a gas with emission lines. Krypton, Xenon, and Argon gases are commonly used, and Xenon and Krypton indeed have emission peaks at 760 and 810 nm.

### 4.3 Injection and recovery of laser pulses

To test the image-difference algorithm and determine detection thresholds, we generated 100 synthetic monochromatic pulses consisting of 2D Gaussians having FWHM determined by the seeing profile of the actual image into which the synthetic monochromatic pulse was injected. The FWHM of the PSF was typically 6 pixels, depending on seeing in each image. We scaled these synthetic monochromatic pulses to various total numbers of photons in the entire profile, from 130 to 800 photons. These synthetic pulses ranged from roughly 0.2x background to 1.5x background photons per pixel. We added these synthetic pulses to actual individual images, simulating a pulse duration less than 1 s, and we placed the pulses at random locations in the image, both in between and coincident with stellar spectra.

Fig. 14 shows a representative synthetic monochromatic laser pulse of sub-second duration having 500 total photons. It was injected into an actual observed raw image. We executed the image-difference analysis to determine if it 'discovered' the synthetic pulses. The fraction of injected pulses that were detected for various total photons in the pulse is shown graphically in Fig. 15. Blindly executing the image-difference algorithm described above, we found the code successfully discovered 50 per cent of the injected pulses that had at least 400 total photons in the profile. It found none of the pulses containing fewer than 300 photons, and it found 100 per cent of the pulses having more than 600 photons.

Thus, the nominal detection threshold at which 50 per cent of the pulses would be detected is 400 photons total within the pulse. This 400 photon threshold represents the number of photons that must be detected in 1 s such that half of such pulses would be detected. The search algorithm has diminishing sensitivity for pulses lasting over 1 s (the exposure time each frame) for which some of the adjacent six bookend exposures would contain the emission, diminishing their contrast with the target image.

This threshold pertains to continuous monochromatic emission. However, the variations in emission-line intensities of ~10 per cent due to seeing and scintillation during 1 sec implies that the continuous laser would need to provide ~4000 photons s⁻¹, on average, in order for the 10 per cent variations to reveal it as a pseudo-pulse. The term 'continuous' here refers to a cadence of pulses (or arrival times of photons) that is more frequent than 1 per second so that our exposure time could not temporally resolve the train of pulses or photons. For example, a train of pulses, each of nanosecond duration and arriving 10⁶ per second would be detected here only as 'continuous' monochromatic emission. Any cadence slower than 1 per s would be detectable as individual pulses, such as from a 'lighthouse' or beacon.

### 5. DISCUSSION AND SUMMARY

We made observations specifically to detect optical light composed of a narrow range of wavelengths, pulsed or continuous, coming from a 10 × 14 degree region towards the centre of the Milky Way Galaxy. We used a special purpose objective prism telescope system that simultaneously surveys a 2 × 3 degree field of view with a fast CMOS sensor that achieves 1 s exposure times and optimizes spectral resolution to identify (and reject) astrophysical objects. A monochromatic light source lasting a few nanoseconds, microseconds, or milliseconds would have been detected in one exposure relative to reference exposures, with a detection threshold of 400 photons in the pulse, corresponding to an incoming 20 000 photons per square meter at the Earth's surface. We obtained 34 800 exposures towards the Galactic Centre region during 2 months in 2021. *We found no monochromatic sources, pulsed or continuous in time, within the 10 × 14 deg region towards the Galactic Centre.*

A major consideration in this optical SETI programme was to minimize the false positives to avoid the time-consuming, and often ambiguous, follow-up effort. We specifically engineered the optics and pixel size to avoid ambiguous false positives, such as from cosmic rays, satellite glints, Cherenkov radiation, or electronics noise that might otherwise require months to interpret (e.g. Sheikh et al. 2021). Towards this end, a key attribute of the optical design was the oversampled point spread function created by the optics and CMOS sensor. Monochromatic point sources illuminate a specific Gaussian-like shape, containing a well-defined ~30 pixels in the core (FWHM ~ 5.5 pixels), which allowed easy identification, and rejection of cosmic rays, Cherenkov radiation, glints off satellites, pulsing lights of airplanes, and electronic noise. None of them would put counts in the pixels consistent with the optical point spread function. Indeed, we found no tantalizing monochromatic signals in 34 800 exposures.

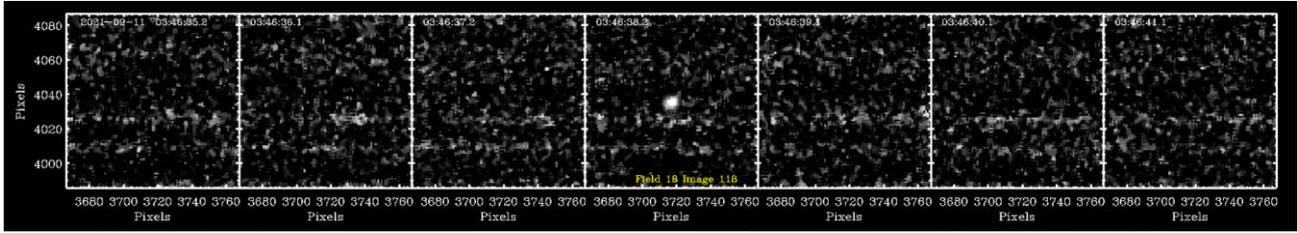

**Figure 14.** A series of 7 consecutive actual images, each 1 s duration, showing a close-up of 100 × 100 pixels. The spectra of two stars are apparent in each frame. A synthetic light pulse of just one wavelength containing 520 photons was injected into the 4th frame. The search algorithm 'discovered' this pulse. This is one representative test case of the detection efficiency for different intensities of light pulse (see Fig. 15). Some synthetic dots are coincident with stellar spectra and others not, and here the injected light pulse resides a few pixels away from a stellar spectrum.

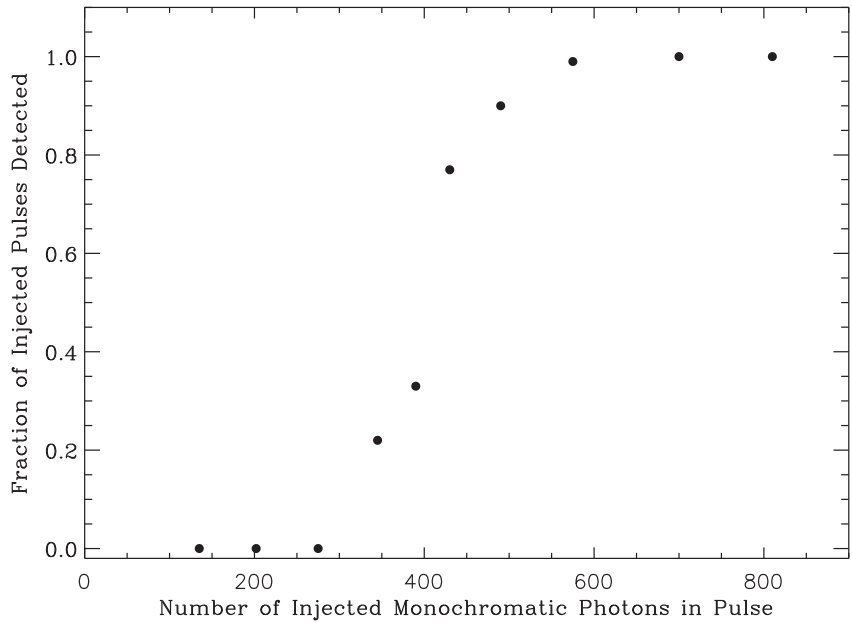

**Figure 15.** The fraction of injected monochromatic pulses detected blindly by the image-difference algorithm as a function of the number of photons in the monochromatic pulse. Pulses containing 400 photons total are detected in 50 per cent of trials. Pulses with 600 photons are detected ~100 per cent of the trials. The nominal detection threshold is 400 photons per pulse.

The detection thresholds for monochromatic pulses of light can be reinterpreted as threshold photon fluxes above the telescope. During a 1 s exposure, our system detects monochromatic pulses containing 400 photons within the 2D PSF, with 50 per cent probability (see Section 4.3). It detects over 99 per cent of the pulses containing > 600 photons. We may translate the 400-photon threshold to a fluence per unit area at the Earth by using the effective collecting area of the 0.278-m RASA telescope system, including efficiency between 450–800 nm and blockage by the camera at prime focus. We find that the effective collecting area is $A_{eff} = 0.020$ m$^2$ (Marcy et al. 2022). Thus, the detection threshold of 400 photons implies a fluence threshold of 20 000 photons per square meter at the Earth's surface for monochromatic pulses of duration less than 1 s.

At wavelengths below 450 nm and above 800 nm the quantum efficiency drops below 50 per cent of peak QE (at ~600 nm), thus requiring more than 20 000 photons per square meter for detection. Atmospheric extinction of several per cent, depending on wavelength, raises this threshold fluence at the top of the Earth's atmosphere by a few per cent, not significant here.

Pulses lasting only a nanosecond or up to 1 s duration would be unresolved in time, i.e. shorter than the 1 s exposure time, and thus could be detected also. Our optics and algorithm perform with the same efficiency independent of the actual duration of the pulse, up to 1 s. Our search included the 10 deg x 14 deg region centred on the Galactic Centre, but the observations were confined in time. By spreading a set of 600 exposures over 10 minutes, and by acquiring 2 or 3 such data sets during 2 months, we covered a range of cadences of monochromatic light sources. Still, some cadences would likely elude our detection, such as one pulse per day, week, or month, with a pulse arriving while we were not observing. Obviously, arbitrary pulse cadences with low duty cycle bring a lower detection probability. More observations are warranted to provide greater cadence coverage. This survey effectively searches for cadences of at least one pulse every 10 minutes. This is just one attempt, with more planned for the future.

One may consider a benchmark laser that is diffraction-limited, with a 10-m diameter aperture, and located 100 ly from Earth. The emitted laser beam has an opening angle of ∼0.01 arcsec for a wavelength of 500 nm. To produce a photon flux at Earth having the detection threshold of 20 000 photons per square meter, a power of 60 MW is required during the 1 s pulse. For a laser launcher located 1000 ly away, a power of 6000 MW is required. Extinction from interstellar dust will increase this requirement by 10–20 per cent. The laser beam footprint at Earth would have a diameter of 0.3–3 au, respectively, a fraction of the area of the inner Solar system. Lasers having apertures smaller than 10 m can also be detected, but their larger opening beam angle would require more laser power, increasing inversely as the square of aperture diameter.

The non-detection here of pulsed and long-lived monochromatic optical light adds to a growing list of similar SETI searches. To date, more than 5000 stars have been searched spectroscopically and individually for monochromatic laser emission (Reines & Marcy 2001; Tellis & Marcy 2017; Marcy 2021; Marcy et al. 2022). In addition, several hundred massive stars of spectral types O, B, A, and F have been searched spectroscopically (Tellis et al., in preparation). Many searches for optical pulses of subsecond duration have been performed, covering over half the sky (e.g. Wright et al. 2001; Howard et al. 2004; Stone et al. 2005; Howard et al. 2007; Maire et al. 2020). No optical laser emission has ever been detected, nor optical pulses having any bandwidth.

This project could have detected both pulsed and continuous lasers. However, continuous lasers, i.e. those with pulses arriving more frequently than 1 pulse per second, probably would have been detected previously. Astronomers have performed many spectroscopic surveys of the entire sky, indeed often using objective prism telescopes, beginning over 100 yr ago (Fleming et al. 1907; Pickering 1912; Cannon & Pickering 1922). Those wide-field spectroscopic surveys revealed thousands of objects that emit emission lines, such as planetary nebulae, H II regions, T Tauri stars, Be stars, Wolf–Rayet stars, M dwarf flare stars, and active galactic nuclei, including at high redshift. Such surveys yield lists of candidate emission-line objects that are invariably pursued with follow-up spectroscopy of modest resolution. Emission lines at non-astrophysical wavelengths would attract attention for further study.

Similarly, astronomers have also performed many surveys of the entire sky using broad-band filters. Continuous laser emission would show up as a source that was bright in one wavelength band (i.e. *B*, *V*, *R*, *I*) but darker in other bands, meriting further observations. Follow-up spectroscopy with moderate resolution would quickly identify emission lines at non-astrophysical wavelengths. Such all-sky surveys reach to roughly 18th magnitude. No continuous technological monochromatic sources, e.g. lasers, were discovered in those past (historic) conventional searches for emission-line objects.

One may wonder if astronomers are careless when performing all-sky surveys for emission-line sources. Perhaps strange emission lines are found at non-astrophysical wavelengths but simply are ignored. This suggestion of carelessness seems unlikely. The discovery of all objects listed two paragraphs above were anomalous at first but were recorded and pursued with further spectroscopy, leading to discovery (e.g. Villarroel & Marcy 2022). Optical lasers would likely have been discovered if they were brighter than 18th magnitude during typical exposure times of ∼10 min. The absence of detected laser emission constitutes an implicit SETI non-detection of sources brighter than 18th mag. Only sources fainter than 18th mag remain.

This non-detection does not necessarily imply that extraterrestrial technology itself is absent in the Milky Way. Other domains of wavelength and pulse cadence merit more observations. Also, the Galaxy may contain optical laser beams that simply occupy too small a volume of the Galaxy, missing the Earth (Forgan 2014). Alternatively, the Galaxy may contain few laser beams, or none. Further, there is a vast parameter space yet to survey, many of which are beyond our current technical capabilities (Wright et al. 2018). To push into new domains, the archived data from surveys can be explored with new algorithms to detect non-astrophysical features that might have been missed by the original science goals.

A large domain of *observable* SETI parameter space has been surveyed by heroic and explicit searches for extraterrestrial technology (e.g. Tarter 2001; Wright et al. 2001; Siemion et al. 2013; Margot et al. 2021; Price et al. 2020). However, SETI parameter space has *unintentionally* been surveyed by all-sky searches for natural, astrophysical objects. Sources were found at wavelengths at which few people expected any sources at all, including the radio, microwave, extreme UV, and gamma rays. Yet, those surveys revealed thousands of sources, many of which were faint, photometrically variable, or contained unexpected emission lines. Those unexpected discoveries implicitly constitute *non-detections* of extraterrestrial technology in the same search domains, but not recorded as such. The optical domain has been especially well surveyed by telescopes for over 100 yr, making the absence of optical SETI signals particularly impressive. The dearth of visible glimmers of technology, mirages aside, leaves us staring into a growing optical SETI desert.

## ACKNOWLEDGEMENTS

This work is dedicated to Franklin Antonio, who provided deep insights, inspiration, and technical innovation to this work and to SETI research at both radio and optical wavelengths. We miss him dearly. This work benefitted from valuable communications with Beatriz Villarroel, Franklin Antonio, John Gertz, Ben Zuckerman, Susan Kegley, Brian Hill, Dan Werthimer, Ariana Paul, Roger Bland, Martin Ward, Lars Mattsson, and Paul Horowitz. We thank the team at Space Laser Awareness for outstanding technical help.

## DATA AVAILABILITY

This paper is based on raw CMOS sensor images obtained with Space Laser Awareness double objective prism telescopes. The 33 000 raw images are 125 MB each, totaling 4.1 TB. They are located on a peripheral disc that is not online. All images are available to the public upon the request of G.M., and a transfer method must be identified.

This paper has been typeset from a TeX/LaTeX file prepared by the author.